\documentclass[%
 aip,
 amsmath,amssymb,
 reprint,%
]{revtex4-1}

\usepackage{graphicx}
\usepackage{dcolumn}
\usepackage{bm}

\usepackage[utf8]{inputenc}
\usepackage[T1]{fontenc}
\usepackage{mathptmx}
\usepackage{etoolbox}

\usepackage{xcolor}

\makeatletter
\def\@email#1#2{%
 \endgroup
 \patchcmd{\titleblock@produce}
  {\frontmatter@RRAPformat}
  {\frontmatter@RRAPformat{\produce@RRAP{*#1\href{mailto:#2}{#2}}}\frontmatter@RRAPformat}
  {}{}
}%
\makeatother
\begin{document}

\preprint{AIP/123-QED}

\title[]{
{\color{black} Bosonic Mpemba effect with non-classical states of light}}
\author{Stefano Longhi}
 \email{stefano.longhi@polimi.it}
 \altaffiliation[Also at ]{IFISC (UIB-CSIC), Instituto de Fisica Interdisciplinar y Sistemas Complejos - Palma de Mallorca, Spain}
\affiliation{ Dipartimento di Fisica, Politecnico di Milano, Piazza L. da Vinci 32, I-20133 Milano, Italy
}%


\date{\today}

\begin{abstract}
The Mpemba effect refers to the surprising observation where, under certain conditions, a far-from-equilibrium state can relax toward equilibrium faster than a state closer to equilibrium. A paradigmatic example is provided by the curious fact that hot water can sometimes freeze faster than cold water. The Mpemba effect has intrigued scientists since long time and has been predicted and observed in a variety of classical and quantum systems. {\color{black} Recently, the search for Mpemba-like effects of purely quantum nature has raised a major interest. Here we predict  the emergence of Mpemba effect  in the quantum optics context  exploiting non-classical states of light. By analyzing the decay dynamics of photon fields in a leaky optical resonator or waveguide, it is demonstrated that bosonic Mpemba effect emerges in the context of the quantum nature of light. Specifically, the relaxation dynamics is strongly influenced by the photon statistics of the initially trapped light field. The Mpemba effect is observed when comparing the decay dynamics} of classical light fields (coherent states) with certain non-classical states, such as Fock states, squeezed states and Schr\"odinger cat states.
\end{abstract}
\maketitle

%


The Mpemba effect (ME) \cite{S1,S2,S2b,S3} describes the counterintuitive  scenario where, under certain conditions, a far-from-equilibrium state can
relax toward equilibrium faster than a state closer to equilibrium. A paradigmatic example, known since long time, is provided by the curious fact that warmer water can sometimes freeze faster than colder water \cite{S1,S2,S2b}. The ME has sparked an intense debate for decades \cite{S3,S3b} and, in spite of being extensively studied in many classical systems \cite{S2b,S3,S3b,S4,S5,S6,S3c,S6b,S6c}, it remains a fascinating and somewhat enigmatic phenomenon that continues to intrigue scientists.
Recently, the ME is being intensely explored in the microscopic world \cite{S7,S8,S9,S9b,S10,S11,S11b,S12,S12b,S13,S14,S14b,S15,S16,S17,fuffa,Busch,noo,cin1,cin2,S17b}, with the prediction and observation of quantum versions of the ME. For example,  in the dynamics of closed many-body systems  an  anomalous relaxation process can occur in
which greater initial symmetry breaking of a subsystem leads to faster symmetry restoration \cite{S11,S12,S12b,S14b}. Such results are stimulating the search for Mpemba-like phenomena of purely quantum origin, which could be of relevance in quantum thermodynamics and could inspire innovative technologies in areas ranging from computing to energy storage \cite{S17b}.
{\color{black} While the ME has been demonstrated in various physical systems, it has largely been overlooked by the photonics community.  Recently, the ME has been predicted to arise in the diffusion 
process of light waves in certain photonic lattices under incoherent dynamics \cite{S18}. Such previous study indicated that, rather surprisingly, tightly-localized light waves can diffuse and reach uniform distribution in the lattice faster than weakly-confined fields, thus providing the first example of ME involving light waves. However, this effect concerning light diffusion is purely classical, prompting the key and intriguing question: do  Mpemba-like effects for quantum light exist, i.e. originating from the quantum nature of light, which is fundamental to modern integrated photonic quantum systems?} \\
In this Letter we predict the emergence of the Mpemba effect in quantum optics context 
originating from the quantum nature of light \cite{book}, {\color{black}which does not have any classical counterpart}. Specifically, we consider the decay dynamics of photon fields from a leaky optical resonator or waveguide \cite{R1,R2,R3,R3b,R4,R5,Haus1,R6,R7,R8} (Fig.1), and show that the relaxation dynamics is strongly affected by  the photon statistics of initially trapped light field. In particular, it is shown that bosonic ME {\color{black} arising from the quantum nature of light is observed when 
comparing the decay dynamics of coherent states (classical states of light)} with certain non-classical states of light, such as Fock states, squeezed states  and Schr\"odinger cat states.\\
A single-mode leaky optical resonator {\color{black} with resonance frequency} $\omega_0$, such as a microcavity S side-coupled to a bus optical waveguide B [Fig.1(a)], can be described by the second-quantization system-bath Hamiltonian $\hat{H}=\hat{H}_S+\hat{H}_B+\hat{H}_{I}$ for the photon field, 
where (see e.g. \cite{R3,R3b,Haus1})
\begin{figure}
\includegraphics[width=8 cm]{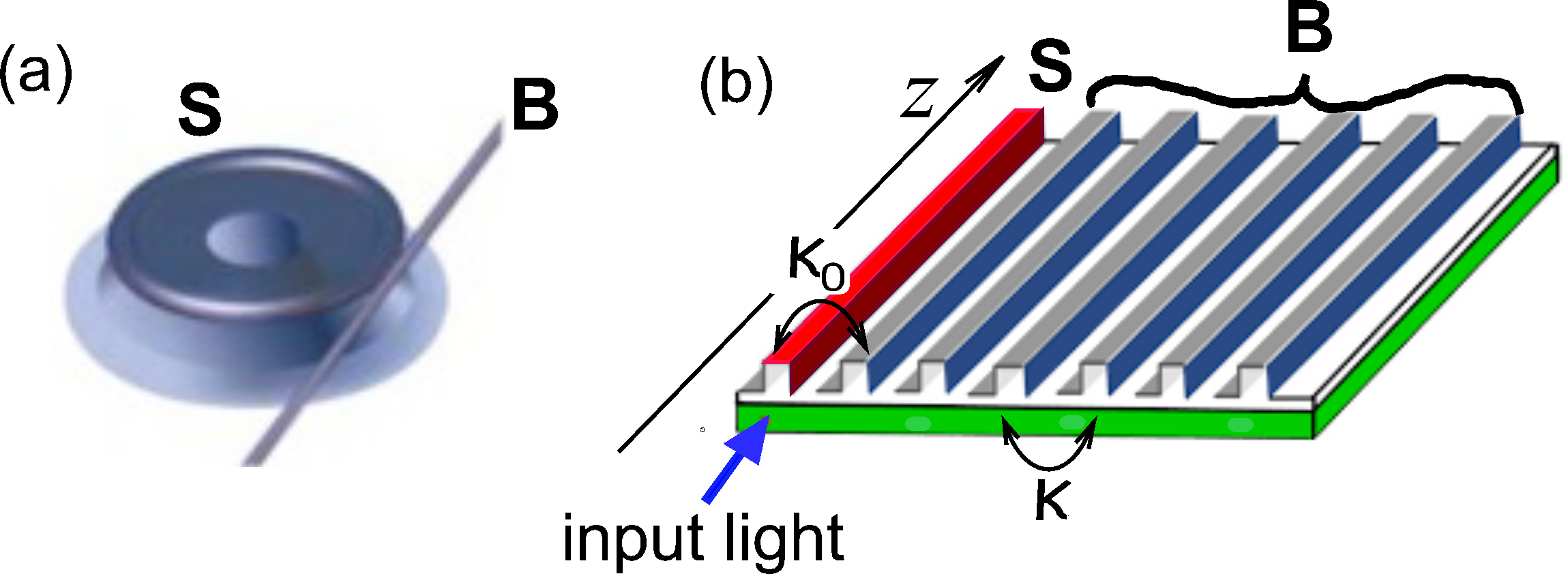}
\caption{Schematic of photon decay in passive optical systems. (a) Optical microcavity S coupled to a bus waveguide B (decay channel). (b) Optical waveguide S side-coupled to a semi-infinite waveguide lattice B.  }
\end{figure}
\begin{equation}
\hat{H}_S=\hbar \omega_0 \hat{a}^{\dag} \hat{a}
\end{equation}
 is the Hamiltonian of the single-mode photon field of the high-$Q$ resonator S, 
\begin{equation}
\hat{H}_B=\sum_{\lambda} \int d \omega \; \hbar  \omega \hat{b}^{\dag}_{\lambda} (\omega)\hat{b}_{\lambda} (\omega)
\end{equation}
 is the Hamiltonian of the photon field in the various decay channels B (propagative or radiation modes) at frequency $\omega$, labelled by the index $\lambda$, and
\begin{equation}
\hat{H}_I= \hbar \sum_{\lambda}  \int d \omega \left\{ V_{\lambda}(\omega) \hat{a}^{\dag} \hat{b}_{\lambda} (\omega)  + {\rm H.c.} \right\}
\end{equation}
describes the coupling between the cavity mode and the decay channel modes. In the above equations, $\hat{a}$, $\hat{a}^{\dag}$ and $\hat{b}_{\lambda} (\omega)$, $\hat{b}^{\dag}_{\lambda} (\omega)$ are the destruction and creation operators of photons in the optical cavity and decay channels, respectively, which satisfy the usual bosonic commutation relations $[\hat{a},\hat{a}^{\dag}]=1$, 
$[\hat{b}_{\lambda}(\omega),\hat{b}_{\lambda'}^{\dag}(\omega')]=\delta_{\lambda,\lambda'}\delta(\omega-\omega')$, $[\hat{a},\hat{a}]=[\hat{b}_{\lambda}(\omega),\hat{b}_{\lambda'}(\omega')]=[\hat{a},\hat{b}_{\lambda}(\omega)]=0$, and $V_{\lambda}(\omega)$ is the spectral coupling function. 
We mention that the above Hamiltonian model also describes the decay of photon fields propagating along the $z$ axis of a leaky single-mode optical waveguide in a linear passive optical network, such as a waveguide S side-coupled to a waveguide lattice B [Fig.1(b)], provided that the replacement $z=ct$ is made \cite{R4,R7,R8}. In this setup, the temporal evolution of the photon field in the leaky optical cavity is mapped onto the spatial propagation of the photon state along the waveguide S. The quantum properties of the photon field in waveguide S of Fig.1(b), or in the resonator S of Fig.1(a), are fully embodied in the reduced density matrix, which can be experimentally retrieved by quantum state tomography methods (see e.g. \cite{H1,H2,H3,H3b,H4,H5,H6,H7} and references therein).\\
 When light is treated at a classical level, {\color{black} indicating by $\alpha(t)=A(t) \exp(-i \omega_0t)$ the electric field amplitude of the resonator mode with carrier (resonance) frequency $\omega_0$ and slowly-decaying complex amplitude $A(t)$}, photon leakage from S is established by the decay law $A(t)$, which satisfies the integro-differential equation (see e.g. \cite{R9,Kurizki0,Kurizki,Kurizki2})
\begin{equation}
\frac{dA}{dt}=-\int_0^t d \tau G(\tau) A(t-\tau)
\end{equation}
where 
\[ G(\tau)=\sum_{\lambda} \int d \omega |V_
{\lambda} (\omega)|^2 \exp [-i(\omega-\omega_0)\tau]\]
 is the reservoir memory function. As is well known, the decay dynamics of $A(t)$ is generally non-exponential owing to memory (non-Markovian) effects \cite{R9,Kurizki,Kurizki2,Facchi1,Facchi2}. However, in the weak-coupling limit and for a non-structured reservoir, the decay becomes exponential and described by the relation  \cite{R9} $|A(t)|^2=|A(0)|^2 \exp(-t /\tau_c)$, where 
$\tau_c=1/(2 \pi \sum_{\lambda} |V_{\lambda}(\omega_0)|^2)$ is the cavity photon lifetime. For example, in the setup of Fig.1(b), indicating by $\kappa_0$ the coupling constant between waveguide S and the adjacent waveguide in the array B, and by $\kappa$ the coupling constant between adjacent waveguides in the array B, the Markovian approximation is attained when $\kappa_0 \ll \kappa$, and the corresponding decay rate is given by $1/\tau_c=(\kappa_0^2 / \kappa)/ \sqrt{1-(\kappa_0 / \kappa)^2}$ \cite{R4,Biagio1,Biagio2}. In the following analysis, we assume rather generally that the decay law $A(t)$ can deviate from an exponential one, however we will assume that $A(t)$ is non-vanishing at any finite time and $\lim_{t \rightarrow \infty} A(t)=0$.\\
At the classical level the initial state is determined solely by the initial electric field amplitude $A(t=0)$ of the resonator mode, and {\color{black} the decay dynamics is the same regardless of the initial field amplitude. 
To unveil the occurrence of ME in the photon decay dynamics, we should consider the quantum nature of light}. To this aim, let us assume that at initial time the photon field is described by the factorized state $|\psi(t=0) \rangle= |\psi_S(0) \rangle_S \otimes |0 \rangle_B$,
where $|0\rangle_B$ is the vacuum state of the zero-temperature reservoir (radiation) modes and
\begin{equation}
| \psi_S(0) \rangle_S=\sum_{n=0}^{\infty} p_n \frac{\hat{a}^{\dag n}}{\sqrt{n ! }} |0 \rangle_S = \sum_{n=0}^{\infty} p_n |n  \rangle_S 
\end{equation}
is the quantum state of the photon field initially {\color{black}{stored in the cavity}}, described by {\color{black} the photon probability amplitudes} $p_n$ in Fock basis $|n \rangle_S \equiv (1/ \sqrt{n !})\hat{a}^{\dag n} |0 \rangle_S$. In the optical setup of Fig.1(b), this initial state is simply accomplished by injecting a light beam in waveguide S with a photon distribution $p_n$. At subsequent times,  the state of photon field evolves according to $|\psi(t) \rangle = \exp(-i \hat{H}t) | \psi(0) \rangle$, and owing to photon leakage the photon field in S becomes highly entangled with the photon field of the reservoir B. The quantum state of light in S is fully described by the reduced density operator
\begin{equation}
\hat{\rho}_S(t)= {\rm Tr}_B {\hat{\rho}}(t)
\end{equation} 
  where 
  \begin{equation}
  \hat{\rho}(t)= |\psi(t) \rangle \langle \psi(t)|=\exp(-i \hat{H} t) | \psi(0) \rangle \langle \psi(0)| \exp(i \hat{H} t)
  \end{equation}
   is the density operator of the full photon field, and the trace in Eq.(6) is taken over the degrees of freedom of the reservoir modes. {\color{black} Since the Hamiltonian is quadratic in the bosonic operators, the exact form of the evolved pure state $|\psi(t) \rangle$, and thus of the density matrix $\hat{\rho}(t)=|\psi(t) \rangle \langle \psi(t)|$, can be readily obtained in the Heisenberg picture (see e.g. \cite{Refere1,Refere2,Valle,Refere3,Referee4}). Once the density matrix of the full photon field has been calculated, the reduced density operator $\hat{\rho}_S(t)$ is obtained  after tracing out the degrees of freedom of the reservoir modes, as shown in Appendix A. Such a procedure is exact and does not require to resort to approximate methods based on Kossakowski-Lindblad-Gorini-Sudarshan or Redfield master equations. 
 The exact form of $\hat{\rho}_S(t)$  in Fock basis reads }
\begin{equation}
\hat{\rho}_S(t)=\sum_{n,m=0}^{\infty} \rho^{(S)}_{n,m} (t) |n \rangle_S \langle m|
\end{equation}   
 where 
 \begin{eqnarray}
 \rho^{(S)}_{n,m} & = & \exp[i(m-n) \omega_0t] \sum_{l=0}^{\infty} \frac{1}{l!} \sqrt{\frac{(l+n)!(l+m)!}{n!m!}} p_{l+n}p_{l+m}^*  \times \nonumber \\
 & \times &A^n A^{* m} (1-|A|^2)^l 
 \end{eqnarray}
 and $A=A(t)$ is the classical decay law, i.e. the solution to Eq.(4) with the initial condition  $A(t=0)=1$. {\color{black} The derivation of Eq.(9) is given in the Appendix A.}
  Since $A(t) \rightarrow 0$ as $t \rightarrow \infty$, for any arbitrary initial {\color{black} probability amplitudes} $\{ p_n \}$ one has $\lim_{t \rightarrow \infty} \hat{\rho}_S(t)=|0\rangle_S \langle 0 | \equiv \hat{\rho}_S^{\infty}$, i.e. as expected the asymptotic  reduced density operator converges toward the vacuum state. {\color{black} The condition $A(t) \neq 0$ at any finite time $t$ ensures that in the transient relaxation process toward the stationary state it is not possible for the system to transiently cross the stationary state, which would introduce some ambiguities in the definition of the ME.}    
 The main point is that the relaxation dynamics strongly depends on the initial probability amplitudes $\{ p_n \}$ of the photon field, which could result in the ME under suitable choice of initial states. {\color{black} To quantify how much the quantum state $\hat{\rho}_S(t)$ is far from $\hat{\rho}_S^{\infty}$ during the relaxation process, a distance measure $D(t)$ for density operators in Hilbert space has to be defined. Such a definition is not unique and ambiguities in the onset of ME could arise \cite{S5,S6c}, however it is generally argued that any distance measure satisfying some minimal properties should
result in a unique definition of the ME \cite{S5}. For quantum systems, the trace distance $D_T(t)$ and the Hilbert-Schmidt distance $D_{HS}(t)$ are generally used as possible metrics of distances in Hilbert space \cite{S9,S9b,N1,N2}. They are defined by the relations}
  \begin{figure*}
\includegraphics[width=18 cm]{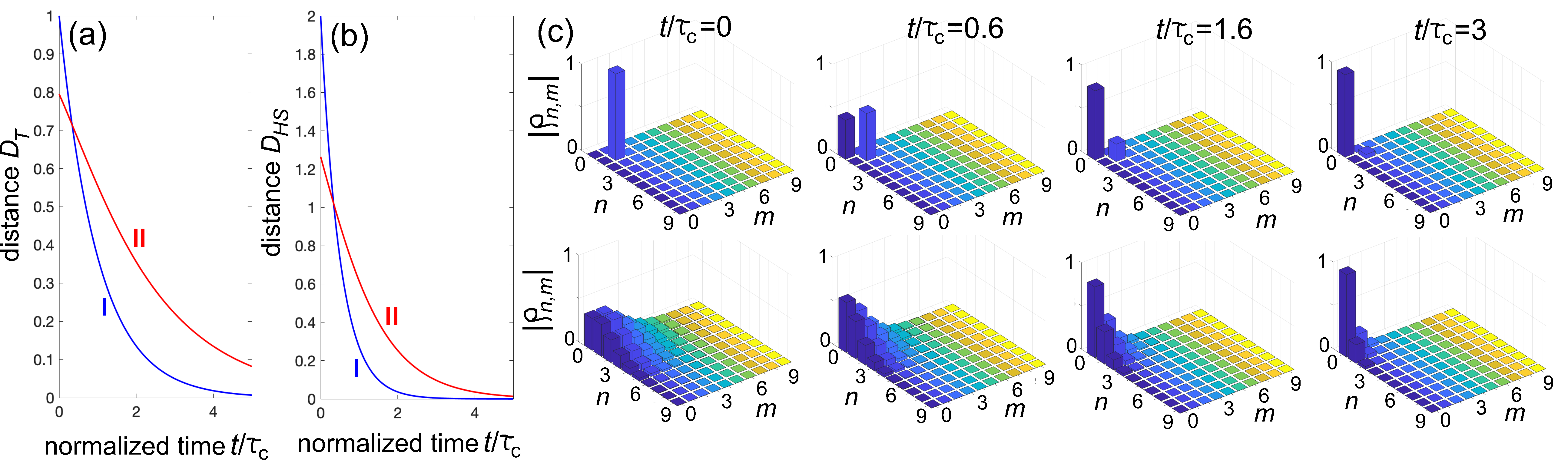}
\caption{Bosonic Mpemba effect in photon decay dynamics with non-classical light. The figure compares the relaxation dynamics in the resonator/waveguide for an initial single-photon Fock state $|1 \rangle_S=\hat{a}^{\dag} |0 \rangle_S$ (state I) and for the coherent state $|\alpha_0^{(coh)} \rangle_S$ (state II) with amplitude $\alpha_0=1$. Panels (a,b)  show the behaviors of trace distance ($D_T$) and Hilbert-Schmidt distance ($D_{HS}$) versus normalized time $t/ \tau_c$. A classical exponential decay law for $|A(t)|^2$ has been assumed, with photon lifetime $\tau_c$. The crossing between the two curves I and II, {\color{black} which arises at about $t/ \tau_c \simeq 0.34$ for both metrics}, is the signature of the ME. Panel (c) shows a few snapshots of the reduced density matrix elements $|\rho^{(S)}_{n,m}(t)|$ at some increasing times [from left to right: $t/ \tau_c=0$ (initial states), $t/ \tau_c=0.6$, $t/ \tau_c=1.6$ and $t/ \tau_c=3$]. Panels in the upper row refer to state I (Fock state decay), whereas panels in the lower row refer to state II (coherent state decay). }
\end{figure*}
 \begin{eqnarray}
D_T(t) & = & \frac{1}{2} {\rm Tr} \left( | \hat{\rho}_S(t)-\hat{\rho}_S^{\infty}| \right) \\
D_{HS}(t) & = & {{\rm Tr} \left( ( \hat{\rho}_S(t)-\hat{\rho}_S^{\infty} )^2 \right)}
\end{eqnarray} 
where $| \hat{O}|$ is the unique positive semidefinite operator defined by $| \hat{O}|= \sqrt{ \hat{O}^{\dag} \hat{O}}$. 
A simple analytical form of the distance $D(t)$ can be derived in two simple cases, namely for a classical-like (coherent) state and for a photon number (Fock) state. When the initial photon distribution is the coherent state $| \alpha_{0}^{(coh)} \rangle_S$ with complex amplitude $\alpha_0$, corresponding to the probability amplitudes $p_n=(1/ \sqrt {n! }) \alpha_0^n \exp(-|\alpha_0|^2/2)$, it can be readily shown that the reduced density operator $\hat{\rho}_S(t)$ is the pure state  $\hat{\rho}_S(t)=| \alpha^{(coh)}(t) \rangle \langle \alpha^{(coh)}(t) |$, where $\alpha^{(coh)}(t)=\alpha_0 A(t)$ describes the classical damping dynamics of the electric field amplitude in S.  In this case one readily obtains
 \begin{equation}
D_{T}(t)= \sqrt{1-\exp(- |\alpha(t)|^2)} \;, \;\; D_{HS}(t)=2 D_T^2(t).
 \end{equation}
The other simple case is provided by the initial state $|\psi_S(t=0) \rangle_S=(1/ {\sqrt N!} ) \hat{a}^{\dag N} |0 \rangle_S=|N \rangle_S$, corresponding to the highly non-classical $N$-photon Fock state.  In this case the reduced density operator is given by the mixed state
\begin{equation}
\hat{\rho}_S(t)= \sum_{n=0}^N  \left(   
\begin{array}{c}
N \\
n
\end{array}
\right) |A(t)|^{2n}(1-A(t)|^2)^{N-n} | n \rangle_S \langle n|.
\end{equation}
Since the reduced density matrix is diagonal in Fock basis, the trace and Hilbert-Schmidt distances can be readily camputed, yielding 
\begin{eqnarray}
D_T(t) & = & 1-\left(1-|A|^2 \right)^N \\
 D_{HS}(t) & = & \sum_{n=1}^N \left\{    
 \left(
 \begin{array}{c}
 N \\
 n
 \end{array}
 \right)|A|^{2n} (1-|A|^2)^{N-n}
 \right\}^2+ \nonumber \\
 & + & \left[ 1-(1-|A|^2)^N \right]^2 .
\end{eqnarray}
For other non-classical states of light, such as  squeezed states and Schr\"odinger cat states \cite{book}, the decay behavior of the distance $D(t)$ can be obtained numerically from Eqs.(10) and (11) using the general form of the density matrix elements $\rho_{n,m}^{(S)}$ in Fock space, given by Eq.(9), after suitable truncation of the density matrix size.

\begin{figure*}
\includegraphics[width=18 cm]{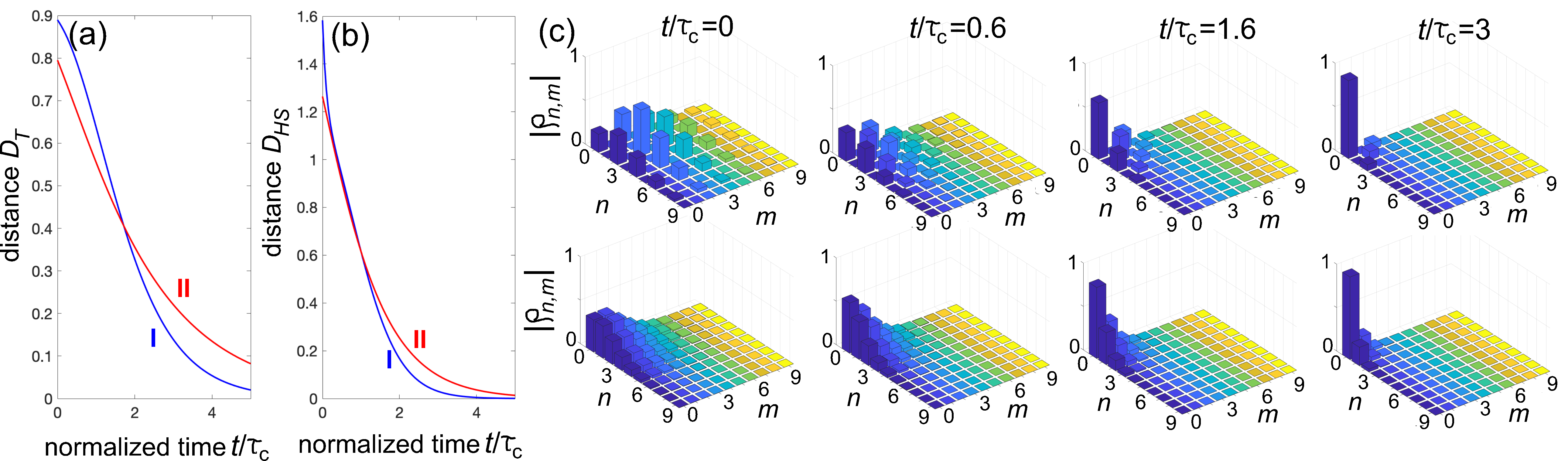}
\caption{Same as Fig.2, except that state I is the Schr\"odinger cat state $|\psi_S(t=0) \rangle_S= \mathcal{N}(|\alpha_0^{(coh)} \rangle_S+ |-\alpha_0^{(coh)} \rangle_S)$ with $\alpha_0=1.5$ ($\mathcal{N}$ is a normalization constant). Panels (a), (b) and (c) display the temporal behaviors of trace distance ($D_T$), Hilbert-Schmidt distance ($D_{HS}$), and the reduced density matrix elements $|\rho^{(S)}_{n,m}(t)|$, respectively. {\color{black} As compared to Fig.2, the crossing time of the two curves I and II depends on the metrics ($t / \tau_c \simeq 1.7$ in (a) and  $t / \tau_c \simeq 1$ in (b)).} }
\end{figure*}

\begin{figure*}
\includegraphics[width=18 cm]{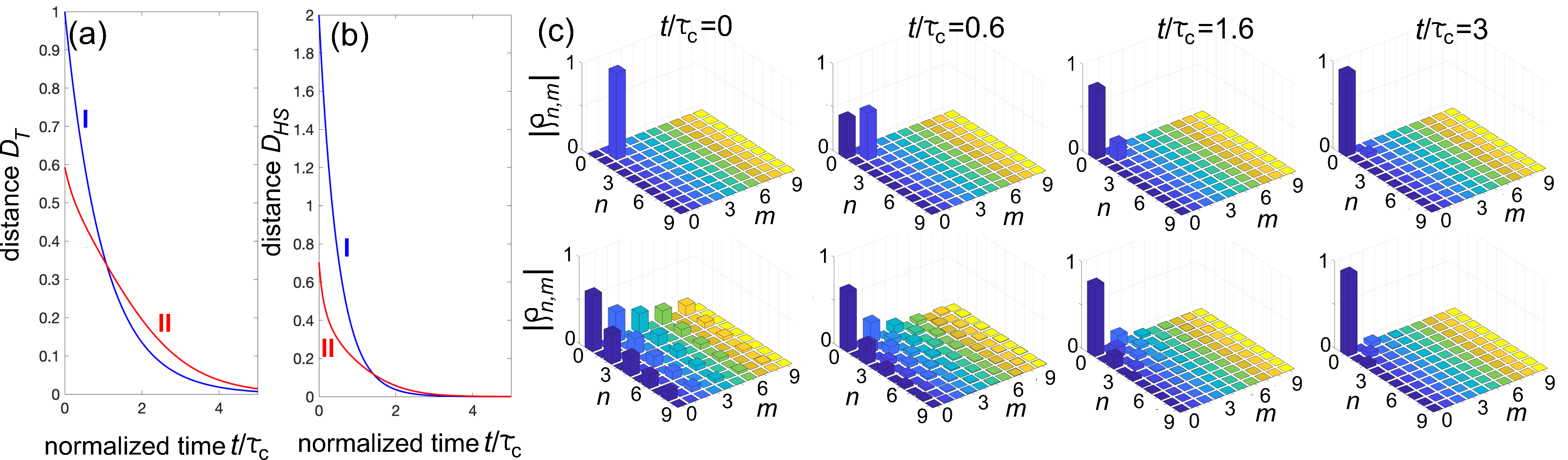}
\caption{Same as Fig.2, except that state II is the squeezed vacuum state $|\psi_S(t=0) \rangle_S= \exp[(1/2)( \xi^{*} \hat{a}^2-\xi \hat{a}^{\dag 2})] |0 \rangle_S$ with squeezing parameter $\xi=1$. Panels (a), (b) and (c) display the temporal behaviors of trace distance ($D_T$), Hilbert-Schmidt distance ($D_{HS}$), and the reduced density matrix elements $|\rho^{(S)}_{n,m}(t)|$, respectively.}
\end{figure*}

{\color{black} To highlight the emergence of the ME in the photon decay process, let us consider two non-equilibrium initial photon states I and II in S at time $t=0$, defined by the two different photon amplitude probabilities $\{ p_{n,I} \}$ and $\{ p_{n,II} \}$. Such two initial states are left at this stages unspecified, with the only constraint that $D^{(I)}(t=0)>D^{(II)}(t=0)$,  i.e. that state I is further from the equilibrium (vacuum) state than state II.} The Mpemba effect arises whenever one has asymptotically $D^{(I)}(t)<D^{(II)}(t)$ as $t \rightarrow \infty$. A general result is that the ME cannot be observed when the initial photon states I and II are coherent states solely, consistently with the classical description of light decay dynamics described above: the decay can deviate from an exponential law due to memory (non-Markovian) effects, however the decay law is the same regardless of the intensity level (mean photon number) initially stored in S. In fact, {\color{black} let us assume that states I and II are two coherent states defined by the classical amplitudes $\alpha_0^{(I)}$ and $\alpha_0^{(II)}$ at time $t=0$, with $| \alpha_0^{(I)}|>|\alpha_0^{(II)}|$. From Eq.(12) it readily follows} that at any time one has  $D^{(I)}(t)>D^{(II)}(t)$, i.e. the initial coherent state with the larger mean number of photons cannot relax toward the vacuum state faster than any other initial coherent state with a smaller mean number of photons. Therefore, to observe the ME non-classical states of light should be considered.
 For example, let us compare the decay laws of the distances $D_T^{(I)}(t)$ for the $N$-photon-number Fock state $|N\rangle_S$ (state I), and $D_T^{(II)}(t)$ for the coherent state $|\alpha_0^{(coh)} \rangle_S$ (state II) , which are given by Eqs.(14) and (12), respectively. Clearly, one has $D_T^{(I)}(0)=1>D_T^{(II)}(0)=\sqrt{1-\exp(-|\alpha_0|^2)}$, i.e. the coherent state is initially closer to the stationary vacuum state, however as $t \rightarrow \infty$ one has  $D_T^{(I)}(t) \sim N |A(t)|^2$ and  $D_T^{(II)}(t) \sim |\alpha_0| |A(t)|$, so that $D_T^{(I)}(t)<D_T^{(II)}(t)$ as $t \rightarrow \infty$, indicating the appearance of the ME. This is illustrated, as an example, in Fig.2, which shows the behaviors of the distances $D^{(I)}(t)$ and $D^{(II)}(t)$ assuming for state I the one-photon Fock state ($N=1$) and for state II a coherent state with initial amplitude $\alpha_0=1$. In the figure, an exponential decay law $|A(t)|^2=\exp(-t /\tau_c)$ has been assumed, with time normalized to the classical photon litefime $\tau_c$. A few snapshots showing the relaxation of the reduced density matrix elements toward equilibrium for the two states I and II are also depicted.{\color{black} We note that non-Markovian (memory) effects, corresponding to deviations of the decay law $|A(t)|^2$ from an exponential curve, would not affect the appearance of the ME effect because they just correspond to a non-linear stretching of the time scale axis}.\\
The ME can be observed when comparing the decay of a coherent state with other non-classical states of light, such as squeezed states or Schr\"odinger cat states, as well when comparing the decay of two different non-classical states. Some illustrative examples  are shown in Figs.3 and 4. Finally, it should be mentioned that to unveil the ME the full properties of the photon field, embodied in the reduced density operator, should be considered: {\color{black}{ the mere observation of some observables, such as the decay of mean photon number $N(t)=\langle \hat{a}^{\dag} \hat{a} \rangle$, is not sufficient to disclose the ME. In fact, for the mean photon number $N(t)$ from Eq.(9) one readily obtains
$N(t) = |A(t)|^2 N(0)$. Hence, if for states I and II, we have $N^{(I)}(0) > N^{(II)}(0)$, then it follows that 
$N^{(I)}(t) > N^{(II)}(t)$ at any subsequent time, and the ME is not observable. However, since $N(t)$ depends only on the diagonal elements of the density matrix, it does not characterize completely the approach to the steady state, which depends on both diagonal and off-diagonal elements of reduced density matrix. Therefore, the distance between the two quantum states in Eqs. (10) and (11) is justified, requiring the ME to demonstrate that a state with a larger initial distance relaxes faster than one with a smaller initial distance.\\
Finally, let us briefly comment on the experimental feasibility to observe the quantum ME using non-classical states of light. }}
The photonic Mpemba effect of genuinely quantum nature predicted in this work should be experimentally observable by considering the propagation of quantum light in integrated quantum photonic circuits, such as in the waveguide array setup shown in Fig.1(b) consisting of a single-mode waveguide S side-coupled to a semi-infinite waveguide lattice, which has been used in previous experiments to demonstrate non-Markovian effects in quantum decay, Zeno-like dynamics and Fano interference using classical and non-classical states of light \cite{R5,Biagio1,Biagio2,Ose}. By considering samples of different length $z$ and using on-chip quantum state tomography  techniques \cite{H6}, the spatial evolution of the density matrix elements   along the propagation distance $z$ can be reconstructed and compared for different input state excitation (such as coherent states and photon number states) of the waveguide S. 
  
In conclusion, we predicted the existence of a bosonic Mpemba effect for light by considering photon decay dynamics of non-classical states of light. The analysis shows that certain non-classical states of light, such as photon number (Fock) states,  decay faster than classical (coherent) light states when the full degrees of freedom of the photon field (i.e. reduced density matrix) are monitored. 
Our results unveil a type of {\color{black} { photonic Mpemba effect of purely quantum nature}}, extending such a counterintuitive and curious phenomenon to the quantum optics context with potential interest in advanced quantum photonic technologies, such as quantum computing, quantum simulations, quantum metrology and quantum communications \cite{fuffa,S17b}, where fast relaxation dynamics could be desirable.  
\begin{acknowledgments}
The author acknowledges the Spanish State Research Agency, through
the Severo Ochoa and Maria de Maeztu Program for Centers 
and Units of Excellence in R\&D (Grant No. MDM-2017-
0711).
\end{acknowledgments}

\section*{Conflict of Interest Statement}
The author has no conflicts to disclose.

\section*{Data Availability Statement}
The data that support the findings of this study are available within the article.

\appendix

{\color{black}
\section{Derivation of the reduced density matrix}
In this Appendix we derive Eq.(9) given in the main text, which provides the general form of the reduced density matrix $\hat{\rho}_S$ in Fock basis when the initial quantum state of the system is 
$|\psi(t=0) \rangle= |\psi_S(0) \rangle_S \otimes |0 \rangle_B$, where $ |\psi_S(0) \rangle_S$ is given by Eq.(5) and $ |0 \rangle_B$  is the vacuum state of the reservoir (radiation) modes.
 To this aim, we first consider the single particle (photon) dynamics, and then derive Eq.(9) for general photon amplitude probabilities  $\{ p_n\}$ of the initial state.\\
\subsection{The single-photon case}
In the single photon case, the initial state of the system is given by $|\psi(t=0) \rangle=  |\psi_S(0) \rangle_S \otimes |0 \rangle_B$, where  $ |\psi_S(0) \rangle_S =\hat{a} |0 \rangle_S=|1 \rangle_S$ is the single-photon Fock state.  Since the dynamics conserves the total number of particles, at subsequent times $t$ the evolved state $|\psi(t) \rangle= \exp(-i \hat{H}t) |\psi(t=0)$ of the photon field can be expanded as (see e.g. \cite{Kurizki0,Valle} )
\begin{equation}
| \psi(t) \rangle=\alpha(t)  \hat{a}^{\dag} |0 \rangle+ \sum_{\lambda } \int {d \omega} \Phi_{\lambda} (\omega,t) \hat{b}^{\dag}_{\lambda}( \omega) |0 \rangle
\end{equation}
 where $|0 \rangle$ is the vacuum state (in both S and B), $\alpha(t)$ is the probability amplitude to find the photon in S at time $t$ (survival amplitude), and $\Phi_{\lambda}(\omega,t)$ is the probability amplitude to find the photon in the reservoir (decay channel) mode $\lambda$ with frequency $\omega$. The probability amplitudes $\alpha(t)$ and $\Phi_{\lambda}(\omega,t)$ satisfy the coupled integro- differential equations (see e.g. \cite{Kurizki0,Valle})
 \begin{eqnarray}
 i \frac{d \alpha}{dt} &= & \omega_0 \alpha(t) + \sum_{\lambda} \int d \omega V_{\lambda}(\omega) \Phi_{\lambda}(\omega,t) \\
 i \frac{d \Phi_{\lambda}}{dt} &= & \omega \Phi_{\lambda}(\omega,t)  + V^*_{\lambda}(\omega) \alpha(t) 
 \end{eqnarray}
 with the initial conditions $\alpha(0)=1$ and $\Phi_{\lambda}(\omega,0)=0$.
 After letting $\alpha(t)=A(t) \exp(-i \omega_0 t)$ and eliminating from the dynamics the amplitudes $\Phi_{\lambda}(\omega,t)$, it turns out that the amplitude probability $A(t)$ satisfies the integro-differential equation (4) given in the main text. To calculate the reduced density matrix, it is worth rewriting Eq.(A1) in the equivalent compact form
 \begin{equation}
 |\psi(t) \rangle= \alpha(t) \hat{a}^{\dag} |0 \rangle+\theta(t) \hat{b}^{\dag}(t) |0 \rangle \
 \end{equation}
where the bosonic creation operator $\hat{b}^{\dag}(t)$ is a linear superposition of the reservoir bosonic operators $\hat{b}_{\lambda}(\omega)$, with $ [ \hat{b}(t), \hat{b}^{\dag}(t) ]=1$, and 
\begin{equation}
| \theta(t)|^2+|\alpha(t)|^2=| \theta(t)|^2+|A(t)|^2=1
\end{equation}
 for particle conservation. In this way, the density matrix $\hat{\rho}(t)=| \psi(t) \rangle \langle \psi(t)|$ of the full system, S+B, can be readily calculated from Eq.(A4) and reads 
\begin{eqnarray}
\hat{\rho}(t) & = & |\alpha(t)|^2 |1 \rangle_S \langle 1| \otimes |0 \rangle_B \langle 0|+ \alpha^*(t) \theta(t) |0 \rangle_S \langle 1| \otimes |1 \rangle_B \langle 0|+ \nonumber \\
& + & \alpha(t) \theta^*(t) |1 \rangle_S \langle 0| \otimes |0 \rangle_B \langle 1|+|\theta(t)|^2 |0 \rangle_S \langle 0| \otimes |1 \rangle_B \langle 1|. \nonumber \\
\end{eqnarray}
where we have set $|1 \rangle_S \equiv \hat{a}^{\dag} |0 \rangle_S$ and $|1 \rangle_B \equiv \hat{b}^{\dag} (t) |0 \rangle_B$. The reduced density matrix $\hat{\rho}_S(t)$ is finally obtained from Eq.(A6) 
after tracing over the degrees of freedom of the reservoir modes, $\hat{\rho}_S(t)= {\rm Tr}_B \hat{\rho}(t)$, and reads
\begin{eqnarray}
\hat{\rho}_S(t) & = & |\alpha(t)|^2 |1 \rangle_S \langle 1| +|\theta(t)|^2 |0 \rangle_S \langle 0| \nonumber \\
& = & |A(t)|^2 |1 \rangle_S \langle 1| +(1-|A(t)|^2 |0 \rangle_S \langle 0|,
\end{eqnarray}
i.e. it describes a mixed state.
It can be readily shown that such a reduced density operator is precisely equivalent to Eqs.(8) and (9) given in the main text  in the single-particle limit ($p_n=1$ for $n=1$, $p_n=0$ otherwise).

\subsection{The multi-photon case}
Let us extend the previous procedure to the multi-photon case. The initial state of the photon field is now assumed to be $| \psi(0)= | \psi_S (0) \rangle_S \otimes |0 \rangle_B$
with $| \psi_S(0) \rangle_S=\sum_{n=0}^{\infty} p_n |n  \rangle_S$, where $p_n$ are the photon probability amplitudes in Fock basis $|n \rangle_S=(1/ \sqrt{n!} ) \hat{a}^{\dag n} |0 \rangle_S$. The evolved state $|\psi(t) \rangle=\exp(-i \hat{H}t) | \psi(0) $ of the full photon field is readily calculated in the Heisenberg picture \cite{Refere1,Referee4}, and basically it is obtained from the state $|\psi(0) \rangle$ after the replacement $\hat{a}^{\dag} \rightarrow \alpha(t) \hat{a}^{\dag}+\theta(t) \hat{b}^{\dag}(t)$ (see e.g. \cite{Valle,Referee4}). This yields 
\begin{equation}
| \psi(t) \rangle =\rangle=\sum_{n=0}^{\infty} \frac{p_n}{\sqrt{n!}} \left( \alpha (t) \hat{a}^{\dag}+ \theta(t) \hat{b}^{\dag}(t) \right)^n |0 \rangle.
\end{equation}
Using the binomial expansion of the $n$-th power, after some straightforward calculation one obtains
\begin{equation}
| \psi(t) \rangle= \sum_{l=0}^{\infty} | \phi_l \rangle_S \otimes |l \rangle_B
\end{equation}
where we have set $|l\rangle_B \equiv (1/ \sqrt{l!} ) \hat{b}^{\dag l}(t) |0 \rangle_B$ and
\begin{equation}
| \phi_l (t) \rangle_S \equiv \sum_{\sigma=0}^{\infty}  \sqrt{\frac{(l+\sigma)!}{l! \sigma !}} p_{l+ \sigma}  
\theta^l \alpha^{\sigma} | \sigma \rangle_S.
\end{equation}
The reduced density operator is obtained from the relation
\begin{equation}
\hat{\rho}_S(t)={\rm Tr}_B \left(  | \psi(t) \rangle \langle \psi(t) | \right)
\end{equation}
where the trace is taken over the degrees of freedom of the reservoir modes. Substitution of Eq.(A9) into Eq.(A11) yields
\begin{equation}
\hat{\rho}_S(t)= \sum_{l=0}^{\infty} | \phi_l(t) \rangle_S \langle \phi_l(t)|.
\end{equation}
The elements of the reduced density operator in the Fock basis $|n \rangle_S$ are given by
\begin{equation}
\rho^{(S)}_{n,m}=  _S\langle n | \hat{\rho}_S(t) | m \rangle_S= \sum_{l=0}^{\infty} \langle n | \phi_l  \rangle_S \langle m | \phi_l \rangle^*_S.
\end{equation}
Substitution of Eq.(A10) into Eq.(A13) and taking into accoung that $\alpha(t)=A(t) \exp(-i \omega_0 t)$ and $|\theta(t)|^2=1-|A(t)|^2$ finally yields Eq.(9) given in the main text. }


\end{document}